\documentclass[a4paper,11pt]{article}
\usepackage{jheppub}

\usepackage{graphicx,cancel,float}
\usepackage{amssymb,amsmath}
\usepackage{textcomp}
\usepackage{bm,times}
\usepackage{epsfig,epstopdf,color,multirow}
\usepackage{slashed}

\def\beqn{\begin{eqnarray}}
\def\eeqn{\end{eqnarray}}
\def\beqs{\begin{subequations}}
\def\eeqs{\end{subequations}}
\def\beq{\begin{equation}}
\def\eeq{\end{equation}}
\def\ba{\begin{array}}
\def\ea{\end{array}}

\def\non{\nonumber\\}

\def\hf{\frac{1}{2}}
\def\[{\left[}
\def\]{\right]}
\def\({\left(}
\def\){\right)}

\def\TeV{\rm TeV}
\def\GeV{\rm GeV}
\def\MeV{\rm MeV}

\def\gU{\rm U}
\def\gSU{\rm SU}

\def\mA{\mathcal{A}}

\def\mE{\mathcal{E}}

\def\mM{\mathcal{M}}

\def\mO{\mathcal{O}}

\def\mR{\mathcal{R}}

\def\mT{\mathcal{T}}

\def\figureautorefname~#1\null{Fig.\,#1\null}
\def\tableautorefname~#1\null{Tab.\,#1\null}

\def\equationautorefname~#1\null{Eq.\,(#1)\null}

\preprint{}
\subheader{\it \today}

\title{The gravitational waves from the collapsing domain walls in the complex singlet model}

\author[a]{~Ning Chen,}
\emailAdd{chenning$\_$symmetry@nankai.edu.cn}
\author[a]{~Tong Li,}
\emailAdd{litong@nankai.edu.cn}
\affiliation[\alpha]{School of Physics, Nankai University, Tianjin 300071, China}
\author[b]{~Yongcheng Wu}
\emailAdd{ycwu@physics.carleton.ca}
\affiliation[b]{Ottawa-Carleton Institute for Physics, \\
Carleton University, 1125 Colonel By Drive, Ottawa, Ontario K1S 5B6, Canada}

\abstract
{\\[1mm]
We study the CP domain walls and the consequent gravitational waves induced by the spontaneous breaking of the CP symmetry in the complex singlet extension to the Standard Model. We impose the constraints from the unitarity, stability and the global minimal of the vacuum solutions on the model parameter space. The CP domain wall profiles and tensions are obtained by numerically solving the relevant field equations.
The explicit CP violation terms are then introduced to the potential as biased terms to make the domain walls unstable and collapse,
The BBN bound on the magnitude of the energy bias is taken into account.
To achieve sufficiently strong gravitational wave signals, the domain wall tension $\sigma$ is required to be at least $\sigma/{\rm TeV}^3 \sim \mathcal{O}(10^3)$. We find that the gravitational wave spectrum can be probed in the future SKA and/or DECIGO programs, when the typical mass scale is at least $\sim \mathcal{O}(10)$ TeV and the explicit CP violation terms are as small as $\mathcal{O}(10^{-29}) - \mathcal{O}(10^{-27}) $.
The gravitational waves from collapsing domain walls thus provide a complementarity to the probe of extremely small CP violation at high-energy scale.

}

\keywords{Beyond Standard Model, Cosmology of Theories beyond the SM, CP violation}
\arxivnumber{arXiv:2004.abcde}

\begin{document}

\maketitle
\setcounter{page}{2}

\newpage


\section{Introduction}
\label{section:intro}

Topological defects, realized as some non-trivial vacuum objects in the early Universe, can arise as a result of the spontaneous breaking of symmetries in new physics beyond the Standard Model (SM)~\cite{Nielsen:1973cs,Kibble:1976sj,Hindmarsh:1994re,Vilenkin:2000jqa}.
A well-known example is the formation of domain walls which occur when a discrete symmetry is spontaneously broken.
Since the discovery of the GW by the LIGO/Virgo collaboration~\cite{Abbott:2016blz,TheLIGOScientific:2017qsa}, it is widely believed that the further probes of various GW signals in different frequencies can provide an unprecedented window to the new physics beyond the SM.
It is likely that the symmetry breaking patterns in new physics models are associated with non-trivial vacuum structure, which therefore leads to various topological defects~\cite{Battye:2011jj,Brawn2011SymmetriesAT,Chatterjee:2018znk,Eto:2018hhg,Eto:2018tnk}.
The GWs from different topological defects during the early evolution of the Universe have been studied in many early literatures as well as recent ones, such as domain walls~\cite{Vilenkin:1984ib,Vachaspati:1984yi,Hiramatsu:2010yz,Hiramatsu:2013qaa,Kitajima:2015nla,Saikawa:2017hiv,Krajewski:2017czs,Zhou:2020ojf} and cosmic strings~\cite{Vilenkin:1984ib,Vachaspati:1984yi,Caldwell:1991jj,Battye:1993jv,Hindmarsh:1994re,Battye:1996pr,Figueroa:2012kw,Cui:2017ufi,Cui:2018rwi,Dror:2019syi,Chang:2019mza}.

Many extensions to the SM involve some high-energy scales $\Lambda_{\rm NP}$ which are typically much higher than the electroweak scale of few hundred GeV.
Correspondingly, the extended matter fields may have too large masses which are even beyond the probes of the future high-energy $pp$ colliders, such as the FCC-hh~\cite{Benedikt:2018csr} or SppC~\cite{CEPC-SPPCStudyGroup:2015csa} with $\sqrt{s}=100\,\TeV$.
Also, the new fields may mix very weakly to the SM fields, e.g. the SM Higgs doublet.
The joint features of heavy masses and weak mixing of the new physics sector lead to the nightmare scenario for the search in terrestrial experiments.
Therefore, it is tempting to ask if there are complementary experiments to probe such a scenario.

In this work, we study the domain wall solutions arising from the complex singlet extension to the Standard Model (cxSM).
This model has been previously studied extensively on the realizations of the strongly first-order electroweak phase transitions~\cite{Barger:2008jx,Jiang:2015cwa,Chiang:2017nmu,Cheng:2018ajh,Grzadkowski:2018nbc,Kanemura:2019kjg,Chen:2019ebq}.
Besides, it turns out that the cxSM can naturally provide the sources of CP violation (CPV) in addition to the CPV phase in the Cabibbo-Kobayashi-Maskawa (CKM) matrix.
The cxSM with the spontaneous CP violation (SCPV) were previously studied in Refs.~\cite{Darvishi:2016gvm,Chao:2017oux,Grzadkowski:2018nbc}.
One intriguing point of this scenario is that the extended scalar sector provides no contribution to the two-loop Barr-Zee diagrams for the electric dipole moments (EDM).
Hence the stringent constraints from the EDM measurements~\cite{Baron:2013eja,Andreev:2018ayy} can be avoided~\cite{Chao:2017oux}.
In our study, we focus on the domain wall solutions due to the broken discrete CP symmetry from the additional complex scalar field $\mathbb{S}$.
It was known that the formation of domain walls is problematic in cosmology~\cite{Zeldovich:1974uw}, since they can quickly dominate the energy densities of radiation and matter.
However, this problem in cosmology can be avoided if the domain walls are unstable and hence collapse before they overclose the Universe~\cite{Vilenkin:1981zs,Gelmini:1988sf,Larsson:1996sp}.
To achieve this possibility, one can consider the scenario where the discrete symmetry is approximate and explicitly broken by the so-called biased terms in the model.
Consequently, stochastic GWs would be produced when the domain walls collapse.
By imposing the constraints to the size of the biased terms in the scalar potential, it is possible to estimate the peak frequencies and the spectra of the GWs.
These GW signatures are completely determined by two parameters of the biased domain wall configurations, namely, the domain wall tension $\sigma$ and the energy difference $\Delta V$ between two shifted potential minima.
It should be noted that the mechanism of GW productions is different from those arising from the phase transitions~\cite{Grojean:2006bp,Dev:2016feu,Balazs:2016tbi,Ivanov:2017dad,Wang:2019pet,Chen:2019ebq,Wang:2020jrd}.
The typical peak frequencies of the GWs from the electroweak phase transition are around $\sim \mO(10^{-4}) - \mO(10^{-1})$ Hz, which will be probed from the future satellite-based interferometers, such as the LISA~\cite{AmaroSeoane:2012km,AmaroSeoane:2012je}, Taiji~\cite{Guo:2018npi}, and Tianqin~\cite{Luo:2015ght} programs.
We find that the typical peak frequencies of the GWs from the collapsing domain walls can be as small as $\sim \mO(10^{-9})$ Hz, with the energy scale of the cxSM in the range of $\sim \mO(10)- \mO(100)$ TeV.
Therefore, one envisions such GW signals to be probed at the future radio telescope of square kilometer arrays (SKA)~\cite{Janssen:2014dka} and the Japanese space GW antenna (DECIGO)~\cite{Kawamura:2011zz} with the latter having wider range of typical frequencies of $\sim \mO(0.1) - \mO(10) $ Hz.

The rest of the paper is the following.
In Sec.~\ref{section:cxSM}, we review the setup of the cxSM, and list the minimal terms required for the SCPV.
The additional complex singlet scalar is assumed to develop a vacuum expectation value (vev).
The unitarity, stability, and the global minimal of the vacuum solutions are imposed to the parameter space of the cxSM.
Afterwards, we obtain the domain wall solutions for the SCPV case by solving the relevant field equations numerically in Sec.~\ref{section:CPDW}.
In Sec.~\ref{section:GW}, we obtain the GW signals by adding the small explicit CPV terms to the cxSM potential, which play the role of biased terms to collapse the possible domain walls.
The size of the biased terms should be sufficiently large so that the domain walls collapsed before the epoch of the big-bang nucleosynthesis (BBN).
By numerical estimation of the domain wall tension, we estimate the peak frequencies and the spectrum of the GW signals, and obtain the related signal-to-noise ratio (SNR) at the future SKA and DECIGO programs.
We find that the GW spectrum can be probed in the future SKA and/or DECIGO programs, when the typical mass scales of the cxSM are $\sim \mO(10)-\mO(100)\,\TeV$.
We summarize our findings in Sec.~\ref{section:summary}.
To facilitate the future studies of the cxSM, we present the stability condition to the general potential of the cxSM in~\autoref{app:stability}.


\section{The complex singlet extension to the SM}
\label{section:cxSM}

The most general scalar potential of the cxSM can be written in the following
\beqn\label{eq:VcxSMfull}
V(\Phi\,, \mathbb{S})&=& \mu^2 | \Phi |^2 +  \lambda | \Phi |^4 + \frac{\delta_2 }{2} |\Phi|^2 | \mathbb{S} |^2  +  \frac{b_2}{2} | \mathbb{S} |^2 +  \frac{d_2}{4} | \mathbb{S} |^4  \non
&+& \Big( \frac{ \delta_1  }{4 } |\Phi |^2 \mathbb{S} +  \frac{ \delta_3   }{4 } |\Phi |^2 \mathbb{S}^2 + c.c. \Big)\non
&+& \Big( a_1 \mathbb{S}  +  \frac{ b_1 }{4} \mathbb{S}^2 +   \frac{ c_1  }{6} \mathbb{S}^3  +  \frac{ c_2 }{6} \mathbb{S} |\mathbb{S}|^2  + \frac{  d_1  }{8} \mathbb{S}^4 + \frac{  d_3 }{8} \mathbb{S}^2 |\mathbb{S}|^2 + c.c. \Big)\,,
\eeqn
where $\Phi$ is the $\gSU(2)_L$ Higgs doublet that breaks the electroweak symmetry.
A global ${\rm U}(1)$ symmetry of $\mathbb{S} \to e^{i\varphi} \mathbb{S}$ can be imposed to eliminate all terms with complex coefficients and only leave the terms in the first line of~\autoref{eq:VcxSMfull}.
In the minimal potential with only the terms respecting the global ${\rm U}(1)$ symmetry, when the complex singlet scalar $\mathbb{S}$ develops a vev, the spontaneous breaking of the global $\gU(1)$ symmetry leads to a massless Nambu-Goldstone boson.
One can thus include explicit ${\rm U}(1)$ symmetry breaking terms of ($\delta_{1,3}$, $a_1$, $b_1$, $c_{1,2}$, $d_{1,3}$), which retain the CP symmetry of $\mathbb{S}\to \mathbb{S}^\ast$ in the potential.
As pointed out in Ref.~\cite{Haber:2012np}, to achieve the SCPV in the theory of one complex scalar field, the global $\gU(1)$ symmetry must be explicitly broken by at least two $\gU(1)$ breaking couplings with different $\gU(1)$ charges.
This is to say one needs to select ${\rm U}(1)$ symmetry breaking terms from at least two groups of parameters among the four groups of ($\delta_1$, $a_1$, $c_2$), ($\delta_3$, $b_1$, $d_3$), $c_1$ and $d_1$.

%

\subsection{The mass spectrum with the SCPV}
\label{sec:cxSMspectrum}

To achieve the SCPV, without the loss of generality, we introduce $b_1$ and $d_1$ terms in addition to the minimal potential.
The SCPV potential we consider is the following
\beqn\label{eq:VcxSM_SCPV}
V(\Phi\,, \mathbb{S})&=& \mu^2 | \Phi |^2 +  \lambda | \Phi |^4 + \frac{\delta_2 }{2} |\Phi|^2 | \mathbb{S} |^2  +  \frac{b_2}{2} | \mathbb{S} |^2 +  \frac{d_2}{4} | \mathbb{S} |^4  \non
&+& \Big(  \frac{ b_1 }{4} \mathbb{S}^2 + \frac{  d_1  }{8} \mathbb{S}^4+ c.c. \Big)\,.
\eeqn
Generically, $(b_1\,, d_1)$ are complex parameters, while all other parameters are real.
To facilitate the discussion of SCPV, we take the notations of $(\Re b_1\,, \Re d_1)$ for real parameters in the potential~\autoref{eq:VcxSM_SCPV} in this section. We shall later incorporate the explicit CP violations as biased terms for the domain wall collapse by taking complex $(b_1\,,d_1)$.

The scalar fields are defined as $\Phi=(0, v+h)^T/\sqrt{2}$ and $\mathbb{S}=(v_s e^{i\alpha} + S+iA)/\sqrt{2}$, where $v$ and $v_s$ are the vevs and $\alpha$ is the CP phase.
In terms of the vevs and the CP phase, the above potential becomes
\beqn
V(v\,,v_s\,,\alpha)&=& \hf \mu^2 v^2 + \frac{\lambda}{4} v^4 + \frac{ \delta_2 }{8 }  v^2 v_s^2 + \frac{b_2}{4}  v_s^2 + \frac{d_2}{ 16} v_s^4 \non
&+& \frac{\Re b_1}{4} \cos(2\alpha)  v_s^2 + \frac{ \Re d_1}{16} \cos(4\alpha) v_s^4  \,.
\eeqn
For the solution with $v_s\neq 0$ and $\alpha\neq 0$, the minimization conditions are given by
\beqs\label{eqs:cxSMmin}
\beqn
\mu^2&=& - \lambda v^2 - \frac{\delta_2}{4} v_s^2  \,,\\
\Re b_1&=& - \Re d_1 \cos( 2\alpha) v_s^2 \,, \\
b_2&=& \frac{\Re d_1 - d_2}{2} v_s^2 - \frac{\delta_2}{2} v^2 \,.
\eeqn
\eeqs

The scalar mass spectrum is obtained as the following $3\times 3$ matrix
\beqs
\begin{eqnarray}
\mM^2&=& \left( \ba{ccc}
\mM_{hh}^2  &  \mM_{hS}^2  & \mM_{hA}^2   \\
\mM_{hS}^2  &  \mM_{SS}^2  & \mM_{SA}^2   \\
\mM_{hA}^2  &  \mM_{SA}^2  & \mM_{AA}^2   \\      \ea  \right) \,,\\
\mM^2_{hh}&=& 2 \lambda v^2 \; ,\\
\mM^2_{SS}&=& \frac{\Re d_1 + d_2}{2} \cos^2 \alpha \,  v_s^2 \; , \\
\mM^2_{hS}&=&\frac{\delta_2}{2}  \cos\alpha v v_s \; , \\
\mM^2_{AA}&=&\frac{ \Re d_1+ d_2 }{2} \sin^2\alpha \, v_s^2 \; , \\
\mM^2_{hA}&=& \frac{\delta_2}{2} \sin\alpha v v_s \; ,\\
\mM^2_{SA}&=& - \frac{3 \Re d_1 - d_2}{4} \sin2\alpha \,  v_s^2  \; .
\end{eqnarray}
\eeqs
In general, one diagonalizes the $3\times 3$ mass spectrum into $\mR^T \mM^2 \mR = \mM_{\rm diag}^2={\rm diag}(m_1^2\,, m_2^2\,, m_3^2)$.
We parametrize the $3\times 3$ orthogonal matrix $\mR$ as below
\beqn\label{eq:CPmixing}
\mR&=& \left(\begin{array}{ccc}
        1 & 0 & 0 \\
        0 & c_3 & s_3 \\
        0 & -s_3 & c_3
    \end{array}\right)\cdot\left(\begin{array}{ccc}
        c_2 & 0 & s_2 \\
        0 & 1 & 0 \\
        -s_2 & 0 & c_2
    \end{array}\right)\cdot\left(\begin{array}{ccc}
        c_1 & s_1 & 0 \\
        -s_1 & c_1 & 0 \\
        0 & 0 & 1
    \end{array}\right) \non
&=&\left(
\ba{ccc}
c_1 c_2 &   s_1 c_2 &  s_2 \\
-( c_1 s_2 s_3 + s_1 c_3 )  &  c_1 c_3 - s_1 s_2 s_3  &   c_2 s_3 \\
 - c_1 s_2 c_3 + s_1 s_3 &  -(  c_1 s_3 + s_1 s_2 c_3 )  &  c_2 c_3 \\
\ea \right)\,,
\eeqn
with the short-handed notations of $s_i \equiv \sin\alpha_i$ and $c_i\equiv \cos\alpha_i$.
Here, $\alpha_1$ represents the mixings between two CP-even scalars, while $\alpha_{2\,,3}$ are two CPV mixing angles.
Correspondingly, the gauge eigenstates of $(h\,,S\,,A)$ are transformed into mass eigenstates of $(h_1\,, h_2\,, h_3)$ by
\beqn
\left( \ba{c} h \\ S   \\  A  \\ \ea  \right)&=&
\mR \cdot
\left( \ba{c} h_1 \\  h_2   \\  h_3  \\ \ea  \right)\,.
\eeqn
Note that the dependence of $\mR$ on these angles does not affect the physical results.

Based on the mass mixing conventions in~\autoref{eq:CPmixing}, the quartic scalar self couplings are related to the scalar masses and mixing angles as below
\beqs\label{eqs:cxSM_SCPV_SelfCoup}
\beqn
\lambda &=& \frac{1}{2 v^2} \sum_i m_i^2 \mR_{i1}^2\,,\\
\delta_2 &=& \frac{2}{v v_s \cos\alpha } \sum_i  m_i^2 \mR_{i1} \mR_{i2}  \,,\\
\Re d_1&=& \frac{1}{ 2 v_s^2 \sin\alpha} \sum_i \Big[  \frac{1}{  \sin\alpha}  m_i^2 \mR_{i3}^2 - \frac{1}{ \cos\alpha} m_i^2 \mR_{i2} \mR_{i3}    \Big] \,,\\
d_2&=& \frac{1}{ 2 v_s^2 \sin\alpha } \sum_i \Big[ \frac{3}{ \sin\alpha }  m_i^2 \mR_{i3}^2  + \frac{1}{  \cos \alpha }  m_i^2 \mR_{i2} \mR_{i3}  \Big]  \,,\\
\tan\alpha&=& \Big(  \sum_i m_i^2 \mR_{i1} \mR_{i3} \Big)/ \Big(  \sum_i m_i^2 \mR_{i1} \mR_{i2}  \Big) \,.
\eeqn
\eeqs
Two constraints can be obtained for three masses of $m_{1\,,2\,,3}$ and three mixing angles of $\alpha_{1\,,2\,,3}$
\beqn\label{eq:CPVrelation}
\tan^2\alpha&=& \frac{ \mM_{AA}^2 }{ \mM_{SS}^2 }  =  \Big(  \frac{ \mM_{hA}^2 }{ \mM_{hS}^2 }   \Big)^2\,.
\eeqn
In practice, we fix three masses and two mixing angles of $\alpha_{1\,,3}$ and solve for $\alpha_2$ and the SCPV mixing angle $\alpha$ numerically by the relation of~\autoref{eq:CPVrelation}.

We summarize all relevant parameters in two bases below
\beqn
\textrm{physical basis}&:& v\,, v_s\,, \ m_{1\,,2\,,3}\,, \alpha_{1\,,2\,,3} \,,\alpha \non
\textrm{generic basis}&:&\mu^2\,, \Re b_1\,, b_2\,,  \lambda\,, \delta_2\,, \Re d_1\,, d_2 \,.
\eeqn
Due to the two constraints for three masses and mixing angles given in~\autoref{eq:CPVrelation}, there are seven free parameters in the physical basis.
Thus, the number of parameters match in two different basis.
In practice, we shall use the parameters in the physical basis and convert them into the parameters in the generic basis by using the relations of~\autoref{eqs:cxSM_SCPV_SelfCoup}.
We  derive the domain wall solutions by using the parameters in the generic basis.

\subsection{The theoretical constraints: unitarity, stability, and the global minimal}
\label{sec:constraints}

Several theoretical constraints should be imposed to the parameter space of the cxSM before we consider the domain wall solutions.

\subsubsection{The perturbative unitarity}

The Lee-Quigg-Thacker unitarity bound~\cite{Lee:1977yc,Lee:1977eg} should be imposed so that the quartic couplings are not too large.
To study the unitarity bound as well as the stability bound, we only need to focus on the quartic terms in the cxSM potential
\beqn\label{eq:VcxSM_quartic}
 V(\Phi,\mathbb{S}) &\sim& \lambda|\Phi|^4 + \frac{\delta_2}{2}|\Phi|^2|\mathbb{S}|^2 + \frac{d_2}{4}|\mathbb{S}|^4 + \frac{\Re d_1}{8} (\mathbb{S}^4   + c.c. )\non
&\sim& \lambda \Big( \hf h^2 + \hf (\pi^0)^2 + \pi^+ \pi^-  \Big)^2 + \frac{\delta_2 }{4} (S^2 + A^2 ) ( \hf h^2 + \hf (\pi^0)^2 + \pi^+ \pi^- ) \non
&+& \frac{d_2}{16 } (S^2 + A^2)^2 + \frac{\Re d_1 }{16  } ( S^4 - 6 S^2 A^2 + A^4 ) \,.
\eeqn
By taking the neutral states of $| \pi^+ \pi^- \rangle $, $\frac{1}{\sqrt{2} } | \pi^0 \pi^0 \rangle$, $\frac{1}{ \sqrt{2} } | h h \rangle$, $\frac{1}{ \sqrt{2} } | SS \rangle$, and $\frac{1}{ \sqrt{2} } | AA \rangle$, as well as $|h\pi^0\rangle$, $|S\pi^0\rangle$, $|hA\rangle$ and $|SA\rangle$ the $s$-wave matrix reads
\begin{align}\label{eq:a0Even_matrix}
a_0^+ =& \frac{1}{16\pi} \left(
\ba{ccccc}
4\lambda & \sqrt{2} \lambda  & \sqrt{2} \lambda & \frac{\delta_2}{2 \sqrt{2} }  & \frac{\delta_2}{2 \sqrt{2} }   \\
\sqrt{2} \lambda & 3 \lambda  & \lambda  & \frac{\delta_2}{4} & \frac{\delta_2}{4}  \\
 \sqrt{2} \lambda & \lambda  & 3 \lambda  & \frac{\delta_2}{4} & \frac{\delta_2}{4}  \\
 \frac{\delta_2}{2 \sqrt{2} }  &  \frac{\delta_2}{4}  & \frac{\delta_2}{4}   &  \frac{3 (\Re d_1 + d_2) }{4}  &   \frac{-3 \Re d_1 + d_2}{4}   \\
 \frac{\delta_2}{2 \sqrt{2} }  & \frac{\delta_2}{4}   & \frac{\delta_2}{4}   &   \frac{-3 \Re d_1 + d_2}{4}  &  \frac{3 ( \Re d_1+ d_2) }{4}  \\
\ea \right)\,,\\
a_0^- = & \frac{1}{16\pi}\textrm{diag}(2\lambda,\frac{\delta_2}{2},\frac{\delta_2}{2},\frac{d_2-3 \Re d_1}{2})\,.
\end{align}
%
Besides, the $s$-wave matrix among the charged states of $| h \pi^\pm \rangle$, $| \pi^0 \pi^\pm \rangle$, $| S \pi^\pm \rangle$, $| A \pi^\pm \rangle$ is
\beqn
a_\pm&=& \frac{1}{16\pi} \textrm{diag}(2\lambda\,, 2\lambda\,, \frac{ \delta_2}{2} \,, \frac{\delta_2 }{2} )\,.
\eeqn
%
The $s$-wave unitarity conditions are imposed such that $|\tilde  a_0^i|\leq 1$ and $|\tilde  a_\pm^i|\leq 1$, with $ \tilde a_0^i$ being all eigenvalues of matrices of $a_0^\pm$ and $a_\pm$ above.
By using the relations in~\autoref{eqs:cxSM_SCPV_SelfCoup}, the perturbative unitarity condition can impose the unitarity bounds to the Higgs boson masses and mixings.

\subsubsection{The stability of the tree-level potential}
\label{sec:stability}

To study the stability bound to the cxSM potential, we still only need to focus on the quartic terms in~\autoref{eq:VcxSM_quartic}\footnote{The stability condition for the full potential in~\autoref{eq:VcxSMfull} is listed in~\autoref{app:stability}.}.
In this case, we parameterize two scalar fields as follows
\beqn
&&    |\Phi| = r\cos\theta \qquad    \mathbb{S}  = r\sin\theta e^{i\phi} \,.
\eeqn
Thus, the quartic terms of the cxSM potential become
\begin{align}
    V(r,\theta,\phi) &= \frac{r^4}{16}\left[ \left( \Re d_1\cos(4\phi)+d_2 \right)\, \sin^4\theta +2\delta_2\,\sin^2\theta\cos^2\theta+  4\lambda\cos^4\theta \right]\nonumber \\
    &= \frac{r^4}{16} [ ( \Re d_1 y + d_2 - 2\delta_2 + 4\lambda)x^2 + 2(\delta_2-4\lambda)x + 4\lambda ] \nonumber \\
    &\equiv \frac{r^4}{16}F(x,y)
\end{align}
%
where $x \equiv \sin^2\theta \in[0,1]$, $y\equiv \cos(4\phi)\in[-1,1]$.
The stability is ensured if $F(x,y)>0,\ \forall x\in[0,1], \forall y\in[-1,1]$.
Therefore, this should be checked for both the bulk regions inside the boundary as well as all corners and edges.

At four corners of $(x,y)=(0,-1)\,,(0,1)\,,(1,-1)\,,(1,1)$, we have
\begin{align}
    &F(0,\pm1) = 4\lambda\nonumber \\
    &F(1,\pm1) = d_2 \pm \Re d_1\,.
\end{align}
Hence, the stability conditions at four corners are
\begin{align}\label{equ:stability_corner}
    \lambda > 0\quad \&\&\quad d_2+ \Re d_1 > 0 \quad \&\&\quad d_2- \Re d_1 > 0 \,.
\end{align}
At four edges, we have $F(x\,,y)$ being
\begin{align}
    \begin{cases}
        F(0,y) &= 4\lambda \\
        F(1,y) &= \Re d_1 y + d_2 \\
        F(x,\pm 1) &= (d_2\pm  \Re d_1 - 2\delta_2 + 4\lambda)x^2 + 2(\delta_2-4\lambda)x+4\lambda
    \end{cases}
\end{align}
Thus we get:
\beqs
\begin{align}
    x=0: ~~   & \lambda>0 \\
    \label{equ:stability_edge2}
    x=1: ~~   &  d_2- \Re d_1 > 0  \quad\&\&\quad d_2 + \Re d_1 > 0 \\
     \label{equ:stability_edge3}
    y=\pm 1:~~  &  2\delta_2\geq d_2\pm  \Re d_1 + 4\lambda ~ || ~ \delta_2 \leq 4\lambda ~ || ~ 3\delta_2\geq d_2\pm  \Re d_1 + 8\lambda ~ || ~ \delta_2^2 < 4\lambda(d_2\pm \Re d_1)\,,
\end{align}
\eeqs
where the first two conditions are the same as those in~\autoref{equ:stability_corner}. For the bulk regions inside the boundaries, since we do not have any extreme point, no condition should be imposed.

In summary, we need to satisfy both the conditions in~\autoref{equ:stability_corner} and \autoref{equ:stability_edge3}.

\subsubsection{The global minimum condition}

The last constraint involves the cosmological evolution of the cxSM, namely, the vacuum that realizes the EWSB should be the lowest one comparing to other vacuum configurations.
In terms of the classical fields, there may be three different configurations for the symmetry breaking:
\beqn
O&:& h\to 0\,,\quad \mathbb{S} \to 0 \,; \non
A&:& h \to 0 \,,\quad  \mathbb{S}  \to \frac{1}{ \sqrt{2} }v_s e^{i\alpha} \,; \non
B&:& h \to v \,,\quad \mathbb{S} \to \frac{1}{ \sqrt{2} } v_s e^{i\alpha}\,,
\eeqn
As the temperature cools down, the symmetry breaking may occur either by one step via $O \to B$, or by two steps via $O \to A\to B$.
The one-step symmetry breaking occurs if the configuration-$B$ is the only possible Higgs potential minimum, and the two-step symmetry breaking occurs if both configure-$A$ and configuration-$B$ coexist as the Higgs potential minimum.
The vacuum configurations of $A$ and $B$ are obtained by solving the following cubic equations
\beqs\label{eqs:VT_highT}
\beqn
A&:&\frac{\partial V}{\partial h }\Big|_{h=0\,, \mathbb{S}=v_s e^{i\alpha }/\sqrt{2} }=0 \,, \frac{\partial V}{\partial \mathbb{S}}\Big|_{h=0\,, \mathbb{S}=v_s e^{i\alpha }/\sqrt{2} }=0\,,\\
B&:& \frac{\partial V}{\partial h}\Big|_{h=v\,, \mathbb{S}=v_s e^{i\alpha }/\sqrt{2}}=0 \,, \frac{\partial V}{\partial \mathbb{S} }\Big|_{h=v\,, \mathbb{S}=v_s e^{i\alpha }/\sqrt{2}}=0 \,.
\eeqn
\eeqs
The numerical solutions are then fed into $V_0(A)$ and $V_0(B)$, and the global minimum condition $V_0(B) \leq V_0(A)$ will be imposed.
The joint constraint from all theoretical constraints will be imposed and displayed below when we discuss the GW signals.


\section{The CP domain walls in the cxSM}
\label{section:CPDW}

\subsection{The domain wall solution}


The CP symmetry is spontaneously broken in the potential in~\autoref{eq:VcxSM_SCPV} and there are subsequent domain wall configurations associated with non-trivial CP phases, called the CP domain walls.
There are two degenerate vacua at $\pm\alpha$ with $\alpha\in [-\pi/2, \pi/2]$ and the potential has a periodicity of $\pi$ along the direction of $\alpha$.
Now we proceed to obtain the CP domain wall solutions.
We work in the Euclidean basis of
\beqn\label{eq:Euclidbasis}
&&(\phi_h(z)\,,\phi_S(z)\,, \phi_A(z) ) \equiv (h(z)\,, S(z)\cos\alpha(z)\,, S(z)\sin\alpha(z) )\,,\non
&& \tan\alpha(z) = \phi_A(z)/\phi_S(z) \,.
\eeqn
The energy density for the CP domain wall is given by
%
\begin{align}
    \mE_{\rm CP}(z) =& \frac{1}{2}(\partial_z \phi_h)^2 + \frac{1}{2}(\partial_z \phi_S)^2 + \frac{1}{2}(\partial_z \phi_A)^2 + V(\phi_h,\phi_Z,\phi_A) \; ,
\end{align}
with the coordinate $z$ being the spatial dimension perpendicular to the domain wall plane.
The potential in this basis is
\begin{align}\label{eq:VEuclid}
    V(\phi_h,\phi_S,\phi_A) =& \frac{\mu^2}{2}\phi_h^2 + \frac{\lambda}{4} \phi_h^4 + \frac{b_2 + \Re b_1 }{4}\phi_S^2 + \frac{b_2-\Re b_1}{4}\phi_A^2 \nonumber \\
    & + \frac{\Re d_1 + d_2}{16}(\phi_S^4 + \phi_A^4) + \frac{d_2 - 3\Re d_1}{8}\phi_S^2\phi_A^2 - V_0 \;,
\end{align}
with $V_0$ being the potential height at the local minimum
%
%
\begin{align}
    V_0 
    &= -\frac{\lambda}{4}v^4 - \frac{\cos(4\alpha)\Re d_1 + d_2}{16}v_s^4 - \frac{\delta_2}{8}v^2v_s^2 \;.
\end{align}
Here, we have used the minimization condition of~\autoref{eqs:cxSMmin} to eliminate the quadratic terms.

The equations of motion (EOM) for $\vec{\phi} \equiv (\phi_h,\phi_S,\phi_A)$ is expressed in the compact form of
\begin{align}
    \label{equ:DW_EOM}
    \frac{d^2}{dz^2}\vec{\phi} = \vec \nabla_{\phi}V(\vec{\phi}) \;,
\end{align}
with $V(\vec \phi)$ given in~\autoref{eq:VEuclid}, and the boundary conditions being
\begin{align}
    \label{equ:DW_BC}
    \vec{\phi}(z=-\infty) &= (v,v_s\cos\alpha,v_s\sin\alpha)\; ,\nonumber \\
    \vec{\phi}(z=+\infty) &= (v,v_s\cos\alpha,-v_s\sin\alpha)\; .
\end{align}
The~\autoref{equ:DW_EOM} can be treated as the EOMs of a particle rolling between two boundaries in a potential of $U = -V$:
\begin{align}
    \frac{d^2}{dt^2}\vec{r} = -\vec \nabla U \Rightarrow \begin{cases}
        \frac{d\vec{v}}{dt} = -\vec \nabla U \;, \\
        \frac{d\vec{r}}{dt} = v \;.
    \end{cases}
\end{align}
%
In one dimension, the solution ($\phi_{1D}$) can be obtained by starting at the minimum of $U$ (maximum of $V$)\footnote{We choose to start at the minimum point, instead of maximum point, of $U$ to avoid the sensitive dependence on the initial condition at one of the boundary when solving the equation numerically.} and revolving to two boundaries by using Runge-Kutta integration.
The initial condition at the minimum point of $U$ is
\begin{align}
    \phi_{1D} &= \phi^{\max}_{1D}\;, \nonumber \\
    \frac{d\phi_{1D}}{dz} &= \sqrt{2|\Delta \overline V|}\;,
\end{align}
where $\phi^{\max}_{1D}$ is the position of the maximum point of $V$ (minimum point of $U$), and $\Delta \overline V$ is the energy difference between the maximum and minimum point of $V$.
When extending to multi-dimension field space, we adapt the path deformation algorithm in Ref.~\cite{Wainwright:2011kj}.
We start from an initial guess of the path linking two local minima, solving the one-dimension problem along the path.
With the one-dimension solution ($\phi_{1D}$) and the path, we then calculate the `force' acting on the `particle' perpendicular to the path
\begin{align}
    N_\perp = \frac{d^2\vec{\phi}}{d\phi_{1D}^2}\left(\frac{d\phi_{1D}}{dz}\right)^2 - \nabla_{\phi}^{\perp}V(\vec{\phi})\;,
\end{align}
where $\phi_{1D}$ is the field coordinate along the path (choose to be from 0 to the length of the path in field space) and $\nabla_{\phi}^{\perp}V(\vec{\phi})$ is perpendicular component of the gradient.
The path in the field space is then deformed according to the above force with two ends fixed until the perpendicular force $N_\perp$ is negligible.
After obtaining the solutions in terms of $(\phi_h\,, \phi_S\,, \phi_A )$, we convert them back into the fields of $(h\,, S\,, \alpha)$ according to~\autoref{eq:Euclidbasis}.

\begin{figure}
\centering
\includegraphics[width=0.8\textwidth]{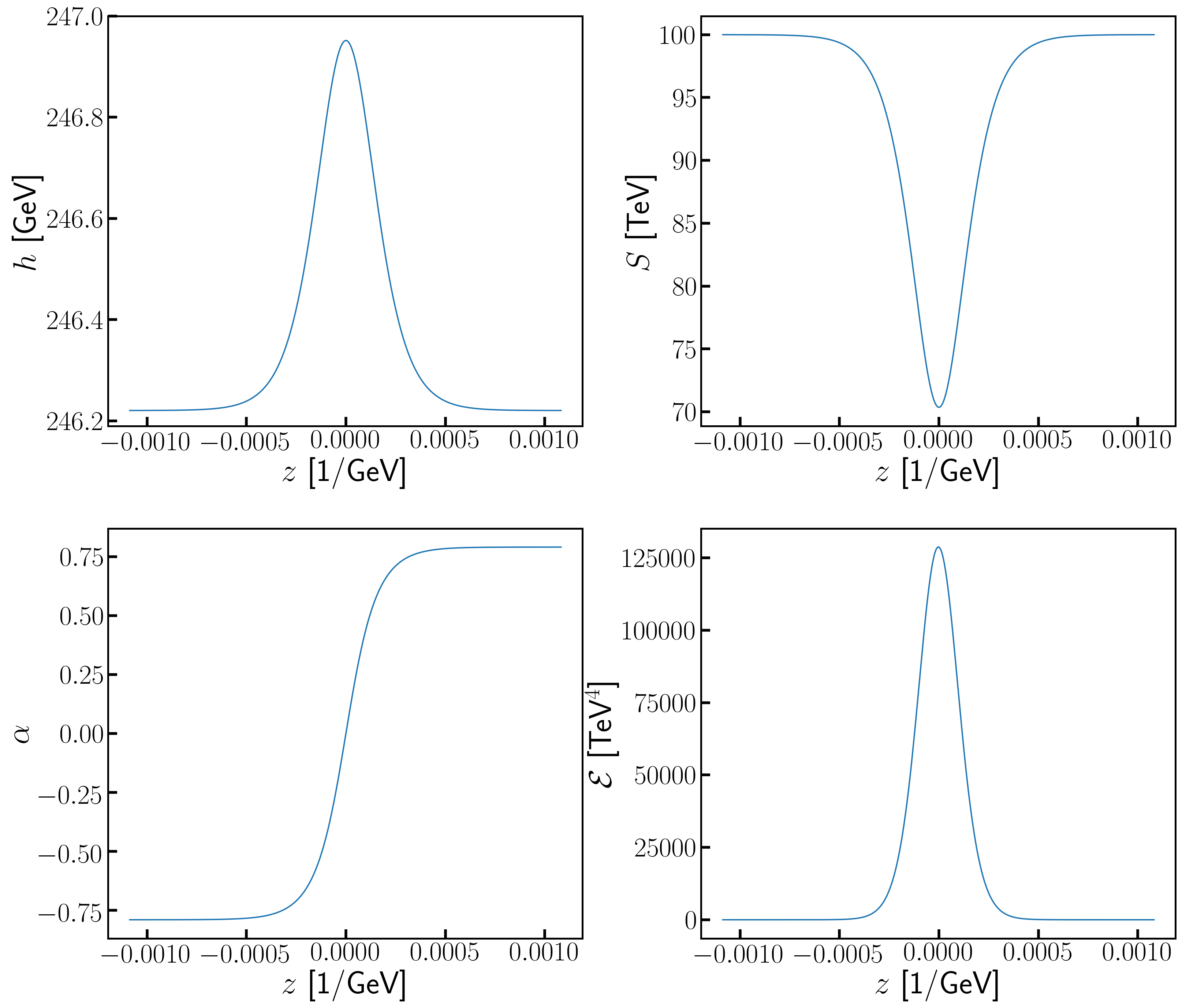}
\caption{
The domain wall profiles of $(h(z)\,,S(z)\,,\alpha(z) )$ and the energy density $\mE(z)$ for the SCPV.
The input cxSM parameters are: $m_1=125\,\GeV$, $m_2=10\,\TeV$, $m_3=10.1\,\TeV$, $v_s=100\,\TeV$, $(\alpha_1\,, \alpha_3)=(10^{-3}\,, 10^{-4}) $.
}
\label{fig:profile}
\end{figure}

In Fig.~\ref{fig:profile}, to illustrate the domain wall solution, we display a domain wall profile in the original basis $(h, S, \alpha)$ and the corresponding energy density $\mE_{\rm CP}(z)$ for the SCPV cxSM.
The input cxSM parameters are $m_1=125\,\GeV$, $m_2=10\,\TeV$, $m_3=10.1\,\TeV$, $v_s=100\,\TeV$, $(\alpha_1\,, \alpha_3)=(10^{-3}\,, 10^{-4})$.
Due to the small mixing angles and small mass difference between $m_2$ and $m_3$, the constraints in~\autoref{eq:CPVrelation} determine that the boundary condition of the CP phase is close to $\alpha\approx \pm \pi/4$.
By integrating the energy density over the direction perpendicular to the wall, we can obtain the tension of domain wall as follows
\beqn\label{eq:DWtension}
\sigma&\equiv& \int_{-\infty}^{+\infty} dz\, \mE_{\rm CP}(z)\,.
\eeqn

\subsection{The biased terms in the cxSM}

As described above, stable domain walls lead to cosmological catastrophy as they overclose the Universe and conflict with the cosmic microwave background (CMB) observed today.
A popular solution is to introduce a small symmetry breaking term, called the biased term, which lifts the degenerate minima.
It makes the walls unstable and annihilate at later times.
In general, the complex $(b_1\,, d_1)$ terms in the above potential~\autoref{eq:VcxSM_SCPV} lead to the explicit CPV as
\beqn
V_{\cancel{\rm CP}}(v\,, v_s\,, \alpha) &=& - \frac{1}{4} \Big(  \Im b_1  \sin(2\alpha) v_s^2 + \frac{\Im d_1}{4}  \sin(4\alpha) v_s^4  \Big) \,,
\eeqn
with $(\Im b_1\,, \Im d_1)$ being the imaginary parts of complex parameters $(b_1\,, d_1)$, respectively.
They are the biased terms for CP domain wall annihilation in the cxSM we consider.
The shift of two degenerate minima induced by the biased potential is then given by
\beqn \label{eq:Vbias}
\Delta V &=& \Big| V_{\cancel{\rm CP}}(v, v_s, \alpha = \frac{1}{2} \cos^{-1} (\frac{ \Re b_1}{-\Re d_1v_s^2}))  - V_{\cancel{\rm CP}}(v, v_s, \alpha = -\frac{1}{2} \cos^{-1} (\frac{ \Re b_1}{-\Re d_1v_s^2})) \Big| \non
 &=& \Big| \frac{ \Im b_1}{2}  \sin(2\alpha) v_s^2 + \frac{ \Im d_1}{8} \sin(4\alpha) v_s^4 \Big|_{\alpha=\frac{1}{2} \cos^{-1} (\frac{\Re b_1}{-\Re d_1 v_s^2})} \,.
\eeqn
With the above biased potential, we can then evaluate the GWs from the domain wall collapse.


\section{The GWs from the collapsing domain wall}
\label{section:GW}

\subsection{The constraints and predictions to the GW signals}

The GWs from the domain wall annihilations were studied in Refs.~\cite{Vilenkin:1981zs,Gelmini:1988sf,Larsson:1996sp,Saikawa:2017hiv}.
With the biased term in the scalar potential, and assuming that the Universe was within the radiation dominated era after the reheating~\footnote{Discussions on the reheating temperature and impacts on the inflation can be found in Ref.~\cite{Gong:2015qha}. }, the temperature when the domain walls annihilated is determined by
\beqn\label{eq:Tann}
T_{\rm ann}&=& 3.41\times 10^{-2}\,{\rm GeV}\, C_{\rm ann}^{-1/2} \mA^{-1/2} \Big(  \frac{ g_{*} (T_{\rm ann})  }{10}  \Big)^{-1/4}  \hat \sigma^{-1/2}  \Delta \hat V^{1/2} \,,
\eeqn
where we have defined the dimensionless quantities of
\beqs
\beqn
\hat \sigma&\equiv&  \frac{\sigma}{1\,\TeV^3} \,,\\
\Delta\hat V&\equiv&  \frac{ \Delta V}{1\,\MeV^4 } \,,
\eeqn
\eeqs
for later convenience.
Here, we take the area parameter as $\mA\simeq 0.8$ for the $\mathbb{Z}_2$ symmetric model.
The $\mathcal{O}(1)$ constant $C_{\rm ann}$ is determined by numerical simulation and is taken to be $C_{\rm ann}=2$ below.
The peak frequency of the GWs at the annihilation time of domain walls is proportional to the annihilation temperature $T_{\rm ann}$, and is given by
\beqn\label{eq:fpeak}
f_{\rm peak}& \simeq & 1.1\times 10^{-7}\,{\rm Hz}\, \Big(  \frac{ g_*( T_{\rm ann} ) }{10}  \Big)^{1/2}  \Big(  \frac{ g_{*s}( T_{\rm ann} ) }{10}  \Big)^{-1/3} \Big(  \frac{ T_{\rm ann} }{1\,\GeV }  \Big)\non
&\simeq&  3.75\times 10^{-9}\,{\rm Hz}\,  \Big(  \frac{ g_*( T_{\rm ann} ) }{10}  \Big)^{1/4} \Big(  \frac{ g_{*s}( T_{\rm ann} ) }{10}  \Big)^{-1/3} C_{\rm ann}^{-1/2} \mA^{-1/2} \hat\sigma^{-1/2}   \Delta \hat V^{1/2}    \,.
\eeqn
Here, $g_*(T_{\rm ann})$ and $g_{*s}(T_{\rm ann})$ count the relativistic degrees of freedom contributing to the energy density and the entropy density, and are both $10.75$ for $1\,\MeV\lesssim T_{\rm ann} \lesssim 100\,\MeV$.
The peak energy density spectrum of the GW is
\beqn
\Omega_{\rm GW}^{\rm peak} h^2(t_0)&=& 7.2\times 10^{-18} \, \tilde \epsilon_{\rm GW} \mA^2 \Big( \frac{ g_{*s}(T_{\rm ann})  }{10}   \Big)^{-4/3}   \Big(  \frac{ T_{\rm ann} }{ 10^{-2}\,\GeV } \Big)^{-4} \hat \sigma^2 \,,
\eeqn
with $\tilde \epsilon_{\rm GW}\simeq 0.7\pm 0.4$ in the scaling regime~\cite{Hiramatsu:2013qaa}.
By using the annihilation temperature in~\autoref{eq:Tann}, the peak energy density spectrum becomes
\beqn
\Omega_{\rm GW}^{\rm peak} h^2(t_0)&=& 5.3\times 10^{-20}\, \tilde \epsilon_{\rm GW} \mA^4 C_{\rm ann}^2 \Big( \frac{ g_{*s}(T_{\rm ann})  }{10}   \Big)^{-4/3}   \Big(  \frac{ g_{*} (T_{\rm ann})  }{10}  \Big) \hat \sigma^4 \Delta \hat V^{-2}  \,.
\eeqn

When the lift of two degenerate vacua is small enough with the approximate discrete symmetry, large scale domain walls are expected to be formed.
The corresponding upper bound to the energy bias reads~\cite{Gelmini:1988sf}
\beqn\label{eq:Vbias_upper}
&& \frac{ \Delta V}{ V_0} < 0.795\,.
\eeqn
In practice, this condition is easily satisfied in the CP domain wall we consider.
On the other hand, the magnitude of the energy bias should be sufficiently large so that the domain walls should have been collapsed before they took over the energy density in the early Universe.
Furthermore, one must also consider whether the walls spoil the standard scenario of BBN when the wall domination occurs after the time of BBN.
The energy constraints at the BBN epoch thus put a stringent constraint to the domain wall lifetime such that $t_{\rm ann}\lesssim 0.01\,{\rm sec}$.
The lower bound on the magnitude of the energy bias in~\autoref{eq:Vbias} is converted to
\beqn\label{eq:Vbias_lower}
\Delta V^{1/4}&\gtrsim& 5.07 \times 10^{-4}\,{\rm GeV}\, C_{\rm ann}^{1/4} \mA^{1/4} \hat\sigma^{1/4}\,.
\eeqn
By combining \autoref{eq:fpeak} and \autoref{eq:Vbias_lower}, we find the lower bound on the peak frequency of the GWs
\begin{eqnarray}
f_{\rm peak}&\gtrsim& 0.964\times 10^{-9}\,{\rm Hz}\, \Big(  \frac{ g_*( T_{\rm ann} ) }{10}  \Big)^{1/4}  \Big(  \frac{ g_{*s}( T_{\rm ann} ) }{10}  \Big)^{-1/3}\;,
\end{eqnarray}
which is independent of both the domain wall tension of $\sigma$ and the energy bias of $\Delta V$, and resides around the most sensitive frequency of SKA accidentally.

The frequency dependencies of the GW energy spectrum are extrapolated based on the numerical simulation and are given by the following rules~\cite{Hiramatsu:2010yz,Hiramatsu:2013qaa}
\beqs
\begin{eqnarray}
\Omega_{\rm GW}h^2(f<f_{\rm peak})&=&\Omega_{\rm GW}^{\rm peak}h^2 (f/f_{\rm peak})^3\;,\\
\Omega_{\rm GW} h^2(f>f_{\rm peak})&=&\Omega_{\rm GW}^{\rm peak}h^2 (f_{\rm peak}/f)\;.
\end{eqnarray}
\eeqs
in the scaling regime.
We determine the discovery prospects of the GW signals by estimating the SNR of~\cite{Caprini:2015zlo}
\beqn\label{eq:GWSNR}
{\rm SNR}&=& \sqrt{ \mT \int_{f_{\rm min} }^{ f_{\rm max}} df\, \Big[ \frac{ \Omega_{\rm GW}(f) h^2  }{ \Omega_{\rm exp}(f) h^2 }  \Big]^2  }\,,
\eeqn
where $\Omega_{\rm exp}(f) h^2$ stands for the experimental sensitivity for the proposed GW programs.
$\mT$ is the mission duration in years for each experiment, and we assume it to be five here.
For illustration below, we take the threshold SNR of $20$ for discovery.

\begin{figure}[!tbp]
  \centering
  \includegraphics[width=0.4\textwidth]{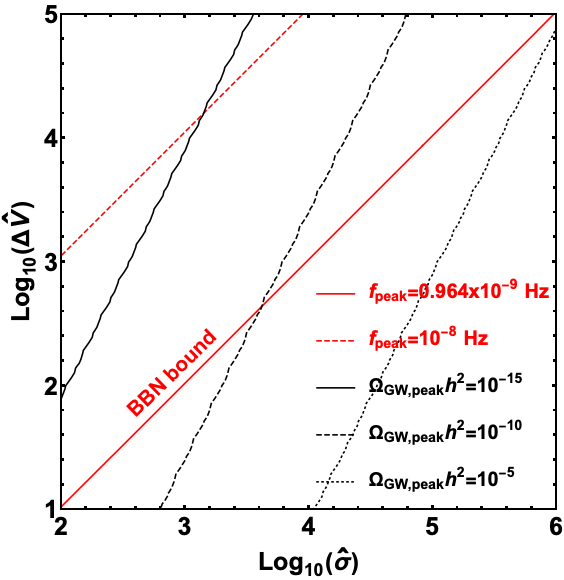}
  \includegraphics[width=0.4\textwidth]{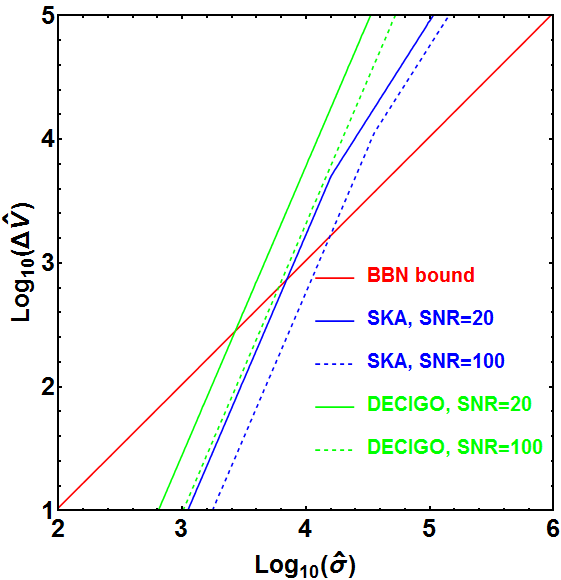}
  \caption{
  Left: The contours of $f_{\rm peak}$ and $\Omega_{\rm GW,peak}h^2$ in the plane of $\log_{10}(\hat \sigma)$ versus $\log_{10}(\Delta \hat V)$.
  Right: The BBN bound (red) and the contours with SNR=20 (solid) and 100 (dashed) for SKA (blue) and DECIGO (green) detectors.
  }
  \label{fig:fOmega}
\end{figure}

In the left panel of Fig.~\ref{fig:fOmega}, we show the model-independent contours of $f_{\rm peak}$ and the GW peak spectrum $\Omega_{\rm GW\,, peak}h^2$ in the plane of $\log_{10}(\Delta \hat V)  $ versus $\log_{10}(\hat \sigma) $.
The BBN lower bound on $\Delta \hat V$ is denoted by the red solid line, which corresponds to the peak frequency of $f_{\rm peak}\simeq 0.964\times 10^{-9}\,{\rm Hz}$.
One can see that, once the $f_{\rm peak}$ is enhanced by one order of magnitude with respect to the BBN bound, the GW peak spectrum decreases more quickly by about eight orders.
Also, along the direction of decreasing $\Delta \hat V$ and increasing $\hat\sigma$, the GW peak spectrum enhances quickly until it hits the BBN bound and exhibits a lower limit on the domain wall tension. From the right panel of Fig.~\ref{fig:fOmega}, we can see that the lower limit on the tension is given by the cross point of the BBN bound and the contour with certain SNR for each GW detector.
We find the minimal tension is $\hat \sigma=7000.9\,(2685.3)$ for SKA (DECIGO) with SNR=20.
Above this limit, for any fixed value of tension, the energy bias $\Delta \hat V$ needs to live in the region between the BBN bound (red curve) and the contour with certain SNR in order to observe GW signal.

\subsection{The future probes of the GWs}

\begin{figure}[!tbp]
  \centering
  \includegraphics[width=0.5\textwidth]{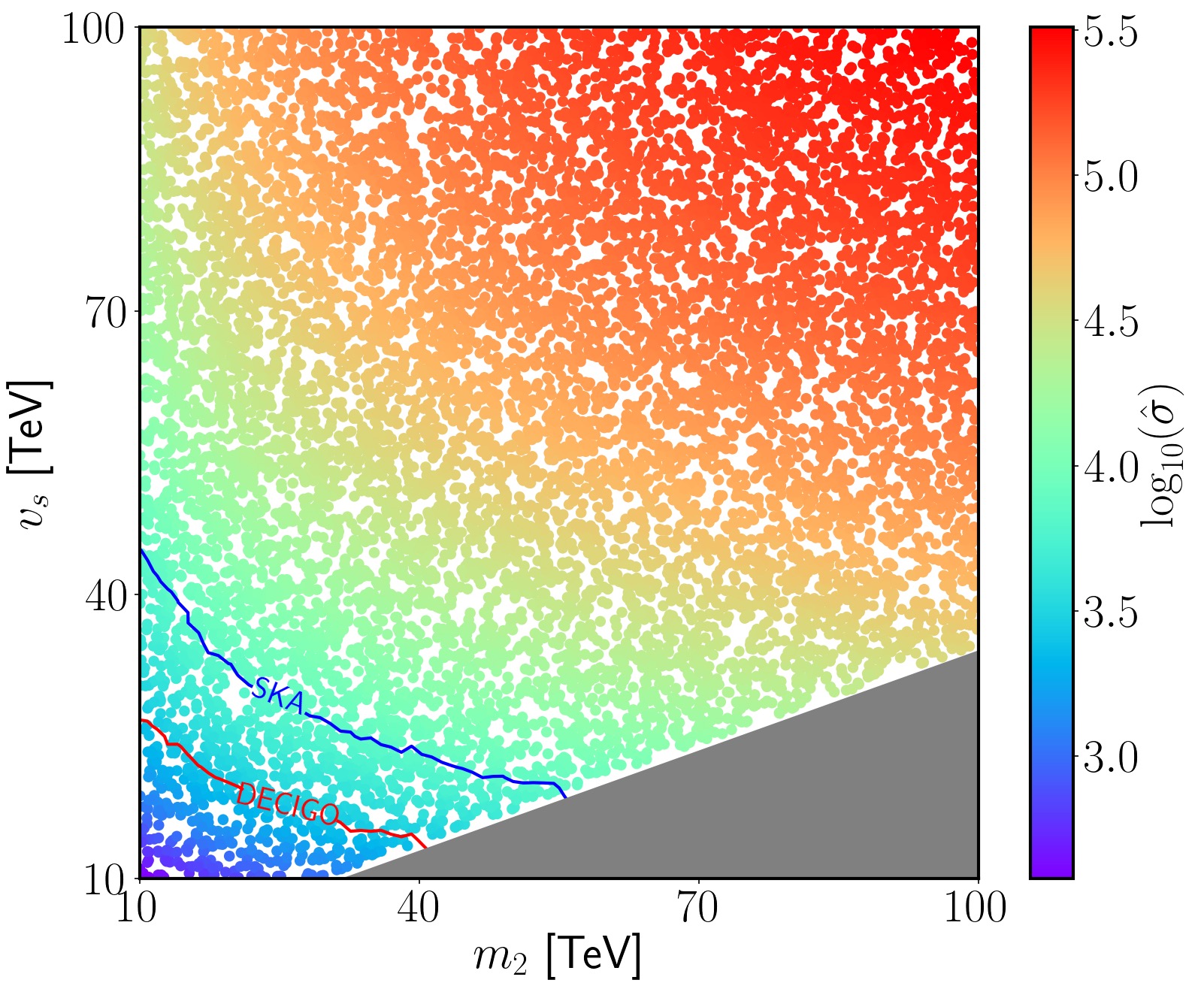}
  \caption{The tension $\log_{10}(\hat \sigma)$ in the plane of $m_2$ versus $v_s$ and the contours for SKA (blue) and DECIGO (red) with SNR$=20$. The grey region has been excluded from the theoretical constraints as presented in~\autoref{sec:constraints}.
  }
  \label{fig:Scan}
\end{figure}

We shall evaluate the GW signals in the future probes of SKA and DECIGO.
We scan the physical parameters in the following ranges
\begin{eqnarray}
10\,{\rm TeV}\leq v_s, m_2\leq 100\,{\rm TeV}\;, m_3=m_2+100\,{\rm GeV}\;.
\end{eqnarray}
In Fig.~\ref{fig:Scan}, we show the domain wall tension $\log_{10}(\hat \sigma)$ in the plane of $m_2$ versus $v_s$.
The grey region has been excluded by the joint theoretical constraints as discussed in Sec.~\ref{section:cxSM}.
The blue and red curves correspond to the minimal tension required by the SKA and DECIGO with $\textrm{SNR}=20$, respectively.
Again, the region above the blue (red) curve leads to sufficiently strong GW signals for the future probes at the SKA (DECIGO) program.

\begin{figure}[!tbp]
  \centering
  \includegraphics[width=0.4\textwidth]{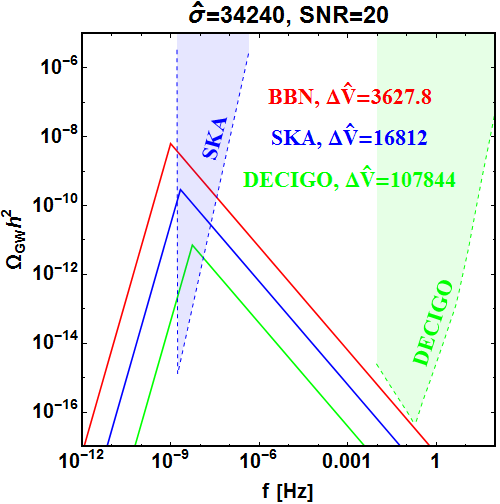}
  \includegraphics[width=0.43\textwidth]{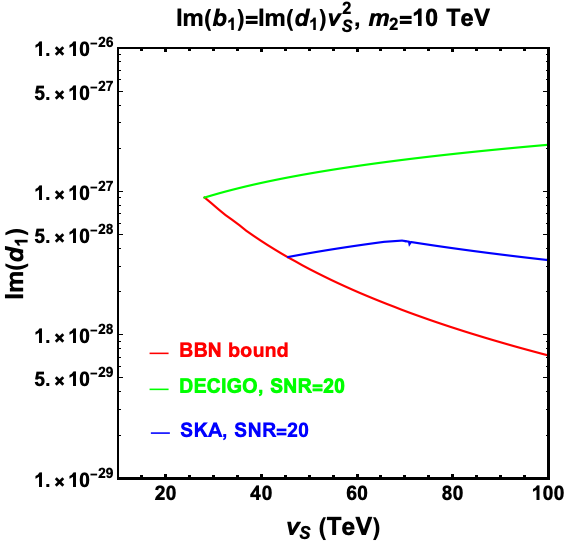}
  \caption{Left: The GW spectra with the bounds of energy bias $\Delta \hat V$ for the benchmark point. Right: The allowed region of $\Im d_1$ as a function of $v_s$. We fix $m_2=10$ TeV and assume $\Im b_1=\Im d_1 v_s^2$.
  }
  \label{fig:Bench}
\end{figure}

To illustrate the probe of the CPV through GWs in the cxSM, we fix the input parameters as
\begin{eqnarray}
&&m_1=125\,{\GeV}\,, m_2=10\,{\TeV}\,, m_3=10.1\,{\TeV}\,, v_s=100\,{\TeV}\,,\non
&&\alpha_1=10^{-3}\,, \alpha_3=10^{-4}\,,
\end{eqnarray}
and obtain the corresponding domain wall tension as $\hat \sigma=34240$. For this benchmark point, based on the BBN bound and the required GW $\textrm{SNR}=20$, one can get the limits on the energy bias as
\begin{eqnarray}
\Delta \hat V &> 3627.8\quad &({\rm BBN})\;,\nonumber\\
\Delta \hat V &< 16812\quad &({\rm SKA\,, SNR>20})\;,\nonumber\\
\Delta \hat V &< 107844\quad &({\rm DECIGO\,, SNR>20}) \;.
\end{eqnarray}
%
The left panel of Fig.~\ref{fig:Bench} (left) displays the GW spectra with the above bounds of energy bias $\Delta \hat V$ for this benchmark point.
Their peak frequencies all reside around $10^{-9}$ Hz.
The peak spectrum is around $10^{-8} (10^{-10}) [10^{-11}]$ for the BBN bound (SKA) [DECIGO].
To observe GW with SNR$>20$, the GW spectra are expected to be lower than the red BBN bound and greater than the blue (green) curve for SKA (DECIGO).
By further varying $v_s\in (10,100)$ TeV and assuming that $\Im b_1=\Im d_1 v_s^2$, in the right panel of Fig.~\ref{fig:Bench}, we convert the bounds of $\Delta \hat V$ to the detectable region of the explicit CPV in the cxSM.
It turns out that the $\Im d_1$ as small as $5\times 10^{-29}-5\times 10^{-28}(10^{-27})$ between the red BBN bound and the upper bound from SKA (DECIGO) with  $\textrm{SNR}>20$ can be probed.

\section{Summary}
\label{section:summary}

In this work, we discuss the possible existence of the topological structures arising from the cxSM.
To testify these structures experimentally, we trace to the relic GWs from the annihilations or decays of these structures.

We consider the spontaneous breaking of the CP symmetry in the cxSM by involving the complex singlet vev and the CP phase.
We impose the constraints from the unitarity, stability and the global minimal of the vacuum solutions on the parameter space of the cxSM.
The joint theoretical constraints exclude the region of $10 \ {\rm TeV}<v_s<40 \ {\rm TeV}$ and $30 \ {\rm TeV}<m_2<100 \ {\rm TeV}$.

The CP domain wall solutions are then obtained by solving the relevant field equations numerically.
The explicit CPV terms are introduced in the potential as biased terms to make the domain walls unstable and collapse.
We consider the BBN bound on the magnitude of the energy bias and find that the lower bound on the model-independent peak frequency of the GWs is around $10^{-9}$ Hz.
By numerical solution of the domain wall tension, we evaluate the peak frequencies and the spectrum of the GW signals, and obtain the related SNR at the future SKA and DECIGO programs.
To achieve sufficiently strong GW signals, for instance with SNR more than 20, the domain wall tension is required to be at least $\hat\sigma \gtrsim \mathcal{O}(10^3)$. Above this minimal tension, the BBN bound and the certain SNR requirement place a constraint on the probed range of the energy bias $\Delta \hat V$.
We find that the GW spectrum can be probed in the future SKA and/or DECIGO programs, when the typical mass scale of the cxSM is at least $\sim \mO(10)\,\TeV$ and the explicit CPV terms are as small as $\mO(10^{-29}) - \mO(10^{-27})$.
The typical energy scales are beyond the scope of the future high-energy $pp$ colliders.
Due to the singlet nature, the CPV mixing in the SM-like Higgs boson cannot be searched for via the future EDM experiments.
The GWs from collapsing domain walls thus provide a complementarity to the probe of extremely small CPV at high-energy scale.


\section*{ACKNOWLEDGMENTS}

We would like to thank Yang Bai, Yin-zhe Ma, Shi Pi, and Kenichi Saikawa for very useful discussions and communication.
The work of NC is partially supported by the National Natural Science Foundation of China (under Grant No. 11575176). TL is supported by the National Natural Science Foundation of China (Grant No. 11975129) and ``the Fundamental Research Funds for the Central Universities'', Nankai University (Grants No. 63191522, 63196013). YW is supported by the Natural Sciences and Engineering Research Council of Canada (NSERC).


\newpage

\appendix

\section{The Stability of the Potential}
\label{app:stability}

In~\autoref{sec:stability}, we have considered a simplified case with $d_3=\delta_3=0$.
 For completeness, here we present the stability condition for the full potential in~\autoref{eq:VcxSMfull}. The relevant quartic terms are
\begin{align}
    V(\Phi,\mathbb{S}) = \lambda|\Phi|^4 + \frac{\delta_2}{2}|\Phi|^2|\mathbb{S}|^2 + \frac{d_2}{4}|\mathbb{S}|^4 + \left(\frac{d_1}{8}\mathbb{S}^4 + \frac{d_3}{8}\mathbb{S}^2|\mathbb{S}|^2 + \frac{\delta_3}{4}|\Phi|^2\mathbb{S}^2 + c.c.\right)
\end{align}
In these terms, $\Phi$ only appears as $|\Phi|^2$, so the SU(2) is preserved. Then the left DOF is $|\Phi|^2\equiv h^2$ and $\mathbb{S} = S + i A$. We parameterized them as
\begin{align}
    h &= r\cos\theta \nonumber \\
    S &= r\sin\theta\cos\phi \nonumber \\
    A &= r\sin\theta\sin\phi
\end{align}
Then the quartic part of the potential become
\begin{align}
    V(r,\theta,\phi) &= \frac{r^4}{16}\left(\sin^4\theta\left(d_1\cos(4\phi)+d_2+d_3\cos(2\phi)\right)+2\sin^2\theta\cos^2\theta(\delta_2+\delta_3\cos(2\phi))+4\lambda\cos^4\theta\right)\nonumber \\
    &= \frac{r^4}{16}\left(x^2(2d_1y^2+(d_3-2\delta_3)y+4\lambda-d_1+d_2-2\delta_2)+x(2\delta_3 y+2(\delta_2-4\lambda))+4\lambda\right)\nonumber \\
    &\equiv \frac{r^4}{16}F(x,y)
\end{align}
where $x = \sin^2\theta \in[0,1]$, $y=\cos(2\phi)\in[-1,1]$. The stability is ensured if $F(x,y)>0,\ \forall x\in[0,1], \forall y\in[-1,1]$. In order to achieve that, we need to check first the corners, then the edges, and last the regions inside the boundaries.

At the four corners ($(x,y)=(0,-1),(0,1),(1,-1),(1,1)$), we have
\begin{align}
    &F(0,-1) = F(0,1) = 4\lambda\nonumber \\
    &F(1,-1) = d_1+d_2-d_3 \nonumber \\
    &F(1,1) = d_1+d_2+d_3
\end{align}
Then the condition are
\begin{align}
    \label{equ:stability_corner_ap}
    \lambda > 0\quad \&\&\quad d_1+d_2-d_3 > 0\quad \&\&\quad d_1+d_2+d_3>0
\end{align}

At the four edges ($x=0$, $x=1$, $y=-1$, $y=1$), we have
\begin{align}
    \begin{cases}
        F(0,y) &= 4\lambda \\
        F(1,y) &= 2d_1 y^2 + d_3 y- d_1 + d_2 \\
        F(x,-1) &= (d_1+d_2-d_3 -2(\delta_2-\delta_3)+4\lambda)x^2 + 2(\delta_2-\delta_3-4\lambda)x + 4\lambda \\
        F(x,1) &= (d_1+d_2+d_3-2(\delta_2+\delta_3)+4\lambda)x^2 + 2(\delta_2+\delta_3-4\lambda)x + 4\lambda
    \end{cases}
\end{align}
Thus we get:
\begin{align}
    \label{equ:stability_edge2_ap}
    x=1: \quad   & d_1 \leq 0 \quad ||\quad \left|\frac{d_3}{4d_1}\right|\geq 1 \quad || \quad d_2-d_1-\frac{d_3^2}{8d_1} > 0\\
    \label{equ:stability_edge3_ap}
    y=\pm1:\quad & d_1+d_2\pm d_3 -2(\delta_2\pm\delta_3)+4\lambda \leq 0 \quad || \quad \delta_2\pm\delta_3-4\lambda\leq 0\nonumber \\
     & ||\quad 3(\delta_2\pm\delta_3)-8\lambda\geq d_1+d_2\pm d_3\quad||\quad  (\delta_2\pm\delta_3)^2 < 4(d_1+d_2\pm d_3)\lambda
\end{align}

Inside the boundaries, we just need to find the minimum point and make sure it is positive. From the form of $F(x,y)$ we find that the extreme points are
\begin{align}
    &\begin{cases}
        x = 0 \\
        y = \frac{4\lambda-\delta_2}{\delta_3}
    \end{cases}\\
    &\begin{cases}
        x = \frac{8d_1(\delta_2-4\lambda)-2\delta_3(d_3-2\delta_3)}{8d_1(d_1-d_2+2\delta_2-4\lambda)+(d_3-2\delta_3)^2}\\
        y = -\frac{2\delta_3(d_1-d_2+\delta_2)+d_3(\delta_2-4\lambda)}{4d_1(\delta_2-4\lambda) + \delta_3(2\delta_3-d_3)}
    \end{cases}
\end{align}
The first one is covered when we consider the edges. Then we just need to make sure either the second extreme point is not inside the boundaries or the value at this point is larger than 0 (no matter it is maximum or minimum). Thus the final condition is

\begin{align}
    \label{equ:stability_region_ap}
    &\frac{8d_1(\delta_2-4\lambda)-2\delta_3(d_3-2\delta_3)}{8d_1(d_1-d_2+2\delta_2-4\lambda)+(d_3-2\delta_3)^2} \leq 0 \quad ||\quad \frac{8d_1(\delta_2-4\lambda)-2\delta_3(d_3-2\delta_3)}{8d_1(d_1-d_2+2\delta_2-4\lambda)+(d_3-2\delta_3)^2} \geq 1 \nonumber \\
    &||\quad \left|\frac{2\delta_3(d_1-d_2+\delta_2)+d_3(\delta_2-4\lambda)}{4d_1(\delta_2-4\lambda) + \delta_3(2\delta_3-d_3)}\right|\geq 1 \quad ||\quad \frac{8d_1^2\lambda + d_1(2\delta_2^2-8d_2\lambda-\delta_3^2)+d_2\delta_3^2+d_3^2\lambda-\delta_2\delta_3d_3}{8d_1(d_1-d_2+2\delta_2-4\lambda)+(d_3-2\delta_3)^2} > 0
\end{align}

In summary, we need to satisfy all the conditions in~\autoref{equ:stability_corner_ap}, \autoref{equ:stability_edge2_ap}, \autoref{equ:stability_edge3_ap} and~\autoref{equ:stability_region_ap}.


\bibliographystyle{JHEP}
\bibliography{references}
\end{document}